\documentclass[aps,pre,preprint,amsmath,amsfonts,showpacs]{revtex4-1}
\usepackage[utf8]{inputenc}
\usepackage[T1]{fontenc}
\usepackage[english]{babel}
\usepackage{amssymb}
\usepackage{graphicx}

\begin{document}
\title{The proton conductivity in benzimidazolium azelate under moderate pressure}
\author{T.~Masłowski$^1$\footnote{t.maslowski@if.uz.zgora.pl}, A.~Drzewiński$^1$, 
P.~Ławniczak$^2$, M.~Zdanowska-Frączek$^2$, J.~Ulner$^3$} 

\affiliation{$^1$Institute of Physics, University of Zielona Góra, ul. Prof. Szafrana 4a, 65-516 Zielona~Góra, Poland\\
$^2$Institute of Molecular Physics, Polish Academy of Sciences, ul. M. Smoluchowskiego 17, 60-179 Poznań, Poland\\
$^3$Institute of Low Temperature and Structure Research PAN, ul. Okólna 2, 50-422 Wrocław, Poland}
\date{\today}

\begin{abstract}
The kinetic Monte Carlo method is applied to examine effects of hydrostatic pressure on the benzimidazolium 
azelate (BenAze) proton conductivity. Following the experimental indications the recently proposed model 
has been modified to simulate the transport phenomena under moderate pressure, resulting in a very good 
agreement between numerical and experimental results. We demonstrate that the pressure-induced changes in 
the proton conductivity can be attributed to solely two parameters: the length of the hydrogen bond and 
the amplitude of lattice vibrations while other processes play a more minor role. It may provide an insight 
into tailoring new materials with improved proton-conducting properties. Furthermore, in high-pressure 
regime we anticipate the crossover from the increasing to decreasing temperature dependence of the proton 
conductivity resulting from the changes in the hydrogen-bond activation barrier with increased pressure.\\
\bigskip

\end{abstract}

\maketitle

\section{Introduction}
Proton conductivity is a very interesting transport phenomenon involved in the long-range charge transfer mechanism. 
The recent great interest in proton transfer unassisted by water comes from the search for fuel cells operating 
above the boiling point of water (373~K) but below extremely high temperatures typical for solid oxides. 
One possibility is to create the proton conductor on the basis of heterocyclic compounds \cite{1,2} that display 
many characteristics to water: they are amphoteric, undergo autoprotolysis and posses the ability to 
form hydrogen bonds. And most importantly, they have substantially higher melting points (e.g.: benzimidazole---447~K). 
Based on these properties various proton conductors, both the polymer \cite{3,4,5,6,7} and crystalline \cite{8,9} 
can be created.
In recent years we have focused on the dicarboxylic acid salts family. Such crystalline anhydrous proton
conductors are organic molecules formed with heterocyclic nitrogen-containing organic chains of imidazole, triazole
and benzimidazole and dicarboxylic acids. These compounds have a number of structural similarities: the chains which 
are formed from acids and heterocyclic molecules are linked by hydrogen bonds while the layers are stabilized by 
the weak electrostatic interaction. In general the proton conductivity increases with temperature according to 
the Arrhenius law which indicates the dominant character of the Grotthuss mechanism \cite{10,11}.

Proton conductivity of anhydrous heterocyclic molecules based materials is a complex process affected by many 
external factors, e.g. temperature and pressure. An external pressure affects the length of the hydrogen bonds 
and modifies the mobility of the structural elements involved in the process. In the literature, the effect of 
pressure on proton conductivity is analyzed mainly for materials exhibiting phase transition to the highly 
conductive phase (superprotonic), for example, CsHSO$_4$ \cite{12} or [(NH$_4)_x$Rb$_{1-x}]_3$H(SO$_4)_2$ \cite{13,14}. 
The application of the hydrostatic pressure causes an increase in the thermal stability of the superprotonic
phase \cite{13}. This is mainly due to an increase in the melting point, while the phase transition temperature varies
slightly with an applied pressure. The similarity of the $p$--$T$ phase diagrams of compounds belonging to the 
$M_m$H$_n(X$O$_4)_{(m+n)/2}$ class of proton conducting crystals (where $M=$ K, Cs, Rb, NH$_4$ and $X=$ S, Se) 
suggests also the possibility of inducing the superionic phase even at the absence of the normal atmospheric pressure
conditions \cite{15,16}.

The electric properties and molecular dynamics of BenAze at ambient pressure have been studied experimentally by 
Zdanowska-Frączek at al. \cite{17,18}. The studies revealed that the temperature dependence of proton conductivity 
can be fitted to the Arrhenius law and the proton conduction is well described by the Grotthuss-type diffusion 
mechanism. Moreover, a detailed description of the proton migration path was proposed \cite{18}. The proton 
conductivity studies performed at different thermodynamic conditions \cite{17,18,19,20} disclosed that BenAze is
extremely sensitive to external pressure. In particular, the proton conductivity values decrease with increasing
pressure and the significant change of the conductivity characteristics shape is observed from linear (an
Arrhenius-type behavior) to nonlinear ones. However, a scaling procedure between the ac conductivity, dc conductivity,
current frequency and pressure (proposed by Sommerfield \cite{21}) leading to one master curve for various temperatures
at ambient pressure is well preserved also for different pressures at a fixed temperature \cite{17}. This indicates
the presence of the same mechanism responsible for proton conductivity at various thermodynamic conditions.

Therefore, we have attempted to support the experimental results observed at elevated pressures by microscopic model
simulations. The theoretical model describing the proton transport in accordance with the Grotthuss mechanism has been
presented recently \cite{22}. For the pure benzimidazole, where the benzimidazole molecules form a long chain connected
to one another by hydrogen bonding, the numerical results were in very good agreement with experimental measurements.
However, the structure of BenAze made of infinite azelaic acid anion chains connected with
benzimidazolium cations by hydrogen bonds is much more complex. Therefore, based on the experimental indications our
model has been modified to reproduce the essential features of BenAze conductivity including the influence of
hydrostatic pressure. 

The theoretical simulations supported by experimental data presented in our paper should provide significant input 
into the understanding of the source of the proton conductivity. Furthermore, some practical conclusions are drawn 
for the high pressure range that has not yet been studied experimentally.

The paper is organized as follows: in Sec. II we present the microscopic model to study effects of moderate
hydrostatic pressure in BenAze. In Sec. III the simulation results are presented and discussed. Finally Sec. IV
concludes our paper, summarizing the main findings.

\section{The proton conductivity model of BenAze}
Recently we have proposed a microscopic model of the proton conductivity based on the kinetic Monte Carlo (KMC) 
approach \cite{22} adequate to characteristic time scales for the proton conduction \cite{23}. The KMC method yields
time evolution of the system provided that all transition rates from every configuration to every other allowed 
state are known \cite{24}. If we count protons crossing a specified position in the chain then we are able to calculate 
the proton current. For the polycrystalline benzimidazole a proton conduction path was modeled by a chain of parallel 
rigid rods whose ends can be occupied by protons. Benzimidazole molecules modeled as rods can rotate by 180$^\circ$ 
flip with or without protons. Protons are also able to migrate by hopping from one rod to the nearest one provided 
the end of the adjacent rod is empty.

It is well established that application of pressure impacts the proton dynamics. The hydrostatic pressure reduces
the atomic volume of a substance and promotes closer packing of atoms (ions) actually modifying the proton transfer
both through an H-bond and by a rotation. The question arises if either of these processes become prevailing when
a moderate hydrostatic pressure is present. According to Fig.~1 there are two noticeable effects of increasing
pressure: the proton conductivity decreases and the range of conductivity values is getting narrower. 
One might suppose that the reduction of rotations at elevated pressure, as a consequence of the increasing activation 
energy, may be responsible for both issues. However, our previous simulations \cite{22} demonstrated that the proton 
current behavior strongly depends on the ratio of the hop frequency to the rotation frequency resulting in two distinct
regimes with significantly different behavior.
As shown, the linear dependence of the proton current on the relative frequency takes place in the 
rotation-dominated regime whereas the current saturates within a broad relative frequency range in the 
hopping-dominated regime. So, if the decrease of the rotation frequency is not accompanied by 
a change in the hopping frequency, both effects of increasing pressure cannot occur simultaneously. 
Therefore, when the pressure-driven modifications of the H-bond potential have turned out to be essential, 
it is worth checking whether the changes in rotations are required to the same extent.

As far as modifications of the H-bond potential are concerned a typical dependence of the pressure on the anionic
group distance is nonlinear \cite{25,26}. So, it can be expected that for a higher pressure the average distance 
between the benzimidazoles decreases slower than linearly. Another very important effect of pressure is a nonlinear 
decrease of the amplitude of lattice vibrations resulting from a damping out the mechanical oscillations by internal 
friction \cite{26,27,28}. When the perturbation theory is employed, the third order elastic constants are required to 
determine the high pressure dependence of the lattice vibrations \cite{26}.

Crucial for our considerations is that BenAze proton transfer path consists of three different types of hydrogen
bridges \cite{17}. Namely, there are O--H$\cdots$O (2.54 and 2.56 \AA), N--H$\cdots$O (2.64 and 2.72 \AA) and
N--H$\cdots$N (2.732 and 2.651 \AA) bonds. It can be assumed that the entire conduction process is limited by
the longest hydrogen bond N--H$\cdots$N. It has the highest H-bond barrier and then the lowest conductance.
Moreover, the energetically favored N--H$\cdots$N angle is close to 180$^\circ$ meaning the H-bond is almost linear
\cite{29} and can successfully be modeled by one-dimensional potential \cite{22}.

Without precise experimental data, we are not able to examine the remaining elements of the transfer path in detail.
But we can propose a simple theoretical model which exploits the existence of the “bottleneck bond”. We use the
black-box concept replacing all other processes (the remaining tunnelings, trans-gauche motion of azelate acid chains
and benzimidazole flips) by a single effective process. Its basic function, as for a rotation in the Grotthuss
mechanism, is to supply a proton to the longest N--H$\cdots$N bond with a certain frequency. In this way, the
cooperative character of the proton transport, the structural reorganization and proton migration along the H-bonds,
has been taken into account.

As the amphoteric nitrogen-based heterocycles demonstrate the presence of both protonated and non-protonated 
nitrogen atoms they can act as donors and acceptors in proton-transfer reactions. For the realistic system the 
concentration of protons on the benzimidazole molecule is slightly shifted from the half-filled case because of various 
crystal structure imperfections. In our simulations the proton concentration is defined by the ratio $c = n/(2N)$, 
where $n$ is the number of protons and $N$ is the number of rods. As there is no indication from the experiment the 
effective concentration was chosen arbitrarily as $c = 0.49$. As we have checked, a change in the effective concentration 
of a few percent does not change the results qualitatively.

We approximate the H-bond potential by the back-to-back Morse functions spanned between the nitrogen anions where
the parameter $d$ represents the distance between them. The larger $d$ the stronger H-bond potential.
\begin{eqnarray}
V_a(x)&=&\frac{1}{2a}\int_{-a}^a \left[V_\text{Morse}\left(\frac{d}{2}-x+y\right)\right. \nonumber \\
&&\left. + V_\text{Morse}\left(x-y-\frac{d}{2}\right)\right] dy\;,\label{va}\\
V_\text{Morse}(x) &=& g \left[\exp\left(-\frac{2x}{b}\right) -2 \exp\left(-\frac{x}{b}\right)\right]\;. \label{pot}
\end{eqnarray}

The hopping between the H-bond minima is defined as the thermally assisted tunneling. In order to model the effect 
of lattice vibrations at a reasonable computational cost we assumed that the position of the nitrogen anions are fuzzy. 
As a consequence, the parameter $a$ responsible for the dispersion in the position of the anions appears in the above 
formula. Due to vibrations the Morse potential barrier is lowered effectively and a current flows more easily. More 
details about the model can be found in  \cite{22}.

\section{Results and discussion}

At ambient pressure the $\sigma_\text{dc}$ conductivity of the BenAze sample \cite{17}, calculated from its bulk
resistance, one may try to approximate by the Arrhenius law. However, at elevated pressure the experimental results
departure significantly from such a behavior (see Fig.~1). 
\begin{figure}[htb]
\centering
\includegraphics[width=8.5cm]{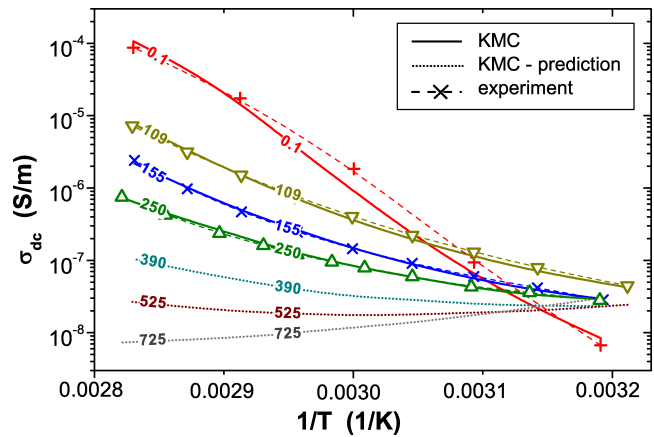}
\caption{The temperature-dependent dc conductivity for BenAze at various pressures. Dashed lines connect the
experimental points: pluses – 0.1 MPa, triangles down – 109 MPa, crosses – 155 MPa, triangles up – 250 MPa. Solid
lines represent KMC simulations, whereas dotted lines come from the KMC predictions at 390, 525 and 725 MPa,
respectively.}
\label{plot01}
\end{figure}
In order to determine the best-fitting parameters the KMC results have been compared simultaneously with available
experimental data apart from the measurements for a pressure of 0.1 MPa. In the latter case the black-box approximation
seems to be too rough because some additional microscopic effects are likely to be significant in the low pressure
regime. The best fit parameter values are presented in Table~\ref{tab1}.

\begin{figure}[htb]
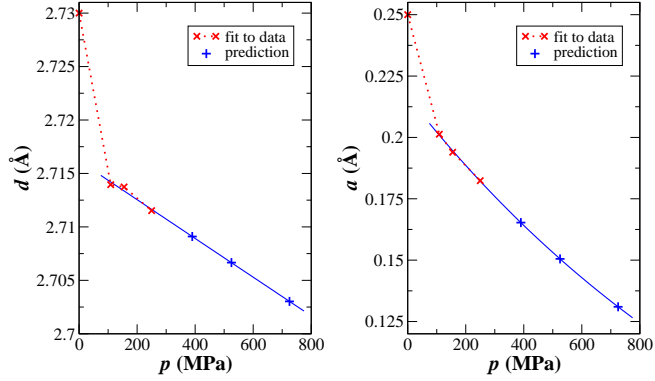

\centering
\parbox{4.24cm}{
\includegraphics[width=4.23cm]{fig02a.eps}
}
\begin{minipage}{4.24cm}
\includegraphics[width=4.23cm]{fig02b.eps}
\end{minipage}
\caption{The crosses represent the H-bond potential parameters, $d$ and $a$, at $T=313$ K obtained from the best fits 
of the dc conductivity data. The dotted lines are a guide for the eye. The pluses denote the same H-bond potential 
parameters extrapolated from the exponential fit to higher pressures at 390, 525 and 725 MPa. The corresponding 
proton conductivities are demonstrated in Fig.~\ref{plot01}.}
\label{plot02}
\end{figure}

\begin{table}[thb]
\caption{Values of model parameters. The definitions of parameters and symbols are the same as in our previous paper \cite{22}.}
\begin{tabular}{@{}*{3}{l}}
\hline
\hline
Parameter  & Symbol  & Value \\
\hline
Frequency of rotation prefactor
& $\nu_R^0$        & $7.5\times10^{7}$ Hz \\
External electric field & $K$ & 0.01 $V/$\AA  \\
Bond length & $d_0$            & 2.73 \AA \\
Thermal expansion coefficient & $d_1$            & 0.0004 \AA$/$K \\
$V_a$ barrier height & $h(T_0)$         & 0.8 eV \\
Distance between minima of $V_a$ & $\Delta x(T_0)$  & 0.713 \AA \\
Reference temperature             & $T_0$            & 313 K \\
$D$ and $L$ defects energy & $V_\text{Coul}$ & 0.2 eV \\
Frequency of hopping prefactor   & $\nu_T^0$        & $10^{9}$ Hz \\
Lattice vibration amplitude        & $a_0$            & 0.25 \AA \\
Thermal susceptibility of $a$    & $a_1$            & 0.0033 \AA$/$K \\
\hline
\hline
\end{tabular}
\label{tab1}
\end{table}
The basic model assumptions have been chosen to satisfactorily reproduce in the KMC simulations the experimental 
data for moderate pressures. This implies a tolerable reproduction of experimental measurements at ambient pressure. 
Such a procedure is dictated by the assumption that because of the loose structure of BenAze the black-box should
work better at higher pressure. In accord with our expectations the $d(p)$ and $a(p)$ functions (the H-bond potential
parameters) have turned out to be nonlinear and monotonic (see Fig.~\ref{plot02}). Moreover, since the damping, as a bulk
phenomenon, weakly depends on the material structure the potential parameter $a$ varies smoothly under pressure. 
In contrast, the pressure compression of bond lengths is more complex and likely only for higher pressure curves 
would become really smooth.

Despite the fact that we have experimental data only for three moderate pressures, we may give some predictions 
about what can be expected for pressures higher than the reported ones \cite{17}. For this purpose, $d(p)$ and
$a(p)$ curves were extrapolated up to 800~MPa using an exponential decay function which should give a qualitatively 
correct fit. The fit included only data for 109, 155 and 250~MPa, as the ambient case could distort the predictions 
because of the loose structure of BenAze. The other model parameters for the elevated pressures are 
the same as in Table~\ref{tab1}.

As the points for experimentally examined pressures 
lies fairly well on the fitted curves for $d(p)$ and $a(p)$ we may expect that the assumptions we made 
should work for some higher pressure. The rotational frequency, $\nu_0^R$, at higher pressure may decrease
as rotations are hindered but the assumption that it may be approximated by a constant value
in a wide range, as it is for 100--250~MPa, should still be valid.
Thus, the predictions presented in Fig.~\ref{plot01} should not be disturbed, at most they could be shifted down.

\begin{figure}[htb]
\centering
\includegraphics[width=8.5cm]{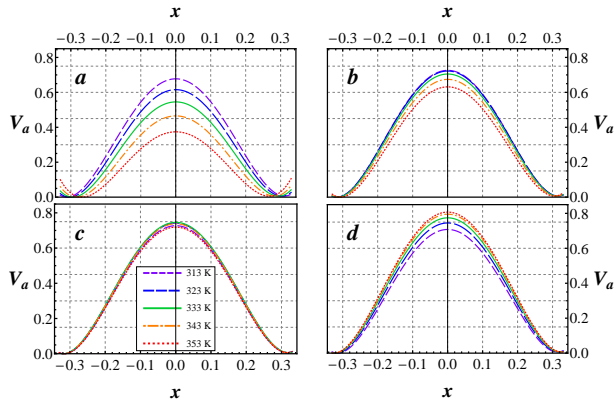}
\caption{The evolution of H-bond potential with increasing pressure plotted between the minima. 
a) A typical behavior for lower pressures, $V_a(x)$ for 109~MPa is given as an example; b) At 390~MPa the inversion of potential curves 
with the temperature ordering starts; c) 525~MPa -- the point where the potential very weakly depends on the temperature; 
d) Above 725~MPa the potential grows with the temperature in the considered range.}
\label{plot03}
\end{figure}

The three exemplary curves in Fig.~\ref{plot01} show the pressure evolution of the thermal conductivity and the crossover 
effect is firmly established. In the interval 500--550~MPa the current becomes almost constant with temperature 
while above this value the proton conductivity, contrary to expectations, 
starts to decrease with increasing 
temperature. For pressure lower than 390~MPa the H-bond barrier decreases with increasing temperature whereas for
higher pressures the behavior of H-bond potential starts to reverse gradually. In the interval 500--550~MPa the shape
of H-bond potential is almost frozen for all examined temperatures. At 725~MPa the process of reversal is finished
and above the H-bond potential barrier increases with increasing temperature.
The explanation may be brought directly from the definition of $V_a(x)$ potential that declines with decreasing $d$ 
but grows with decreasing $a$. Thus, the conductivity behavior with increasing pressure is a result of the interplay of $d$ and $a$. 
It occurs that for BenAze the decline of $a$ wins with the effect of decreasing $d$ resulting in behavior shown in Fig.~\ref{plot03} 
and further in Fig.~\ref{plot01}.
Also, the evolution of thermal behavior of hydrogen bonding can be based on the analysis of the
thermal dependence of $d$ and $a$ parameters. We consider the linear approximation, i.e.:
\begin{eqnarray}
d(T)&=&d_0 + d_1\,(T-T_0)\;,\\
a(T)&=&a_0 + a_1\,(T-T_0)
\label{eeqq}
\end{eqnarray}
and assume that the both linear coefficients, $d_1$ and $a_1$, retain the mutual relationship of the potential
parameters shown in Fig.~\ref{plot02}. 
Naturally, the values of $d_0$, $d_1$, $a_0$ and $a_1$ must be in the proper proportions
to give the behavior presented in Fig.~\ref{plot01}. Otherwise, for example, the conductivity curves could be concave instead
of convex.

In general, the proton conductance of various materials can change substantially with temperature and pressure. At
the level of the microscopic description, as shown by our discussion, the crucial point is the interplay between only
a few parameters: bond length $d_0$, thermal expansion coefficient $d_1$, lattice vibration amplitude $a_0$ and
lattice vibration thermal susceptibility $a_1$. This may serve as an indication of promising compounds among those
exhibiting a low proton conductivity at ambient pressure. Some of them can manifest as a considerable increase in the
conductivity at higher pressures related to the lowering of the activation barrier.
We hope that this knowledge will help technologists to successfully design new materials to be used in electrochemical devices.

\section{Summary}
BenAze belongs to the wide family of organic acids with heterocyclic molecules. Its conductivity is too low for 
practical applications but in view of the well-known crystalline structure it can serve as an excellent system for 
testing. The proton transport efficiency in BenAze is controlled by the longest N--H$\cdots$N hydrogen bonds playing
the role of bottlenecks. In our simulations the proton migration along this H-bond was analyzed thoroughly while
all other transport sub-processes (rotations, tunnelings) were replaced by a single effective process with a fixed 
frequency. The simulation results were found to be in good agreement with the experimental data indicating that the 
basic model assumptions have been chosen properly in a wide range of thermodynamic parameters.

The present simulations confirm a sequence specificity of the proton motion via the Grotthuss mechanism. The
pressure-induced changes in the proton conductivity have occurred to be attributed to solely two parameters:
the length of the  hydrogen bond and the amplitude of lattice vibrations indicating the dominant role of hydrogen
bonds in the proton transport under moderate pressure. It agrees with the hypothesis that in the proton conductors
exhibiting a Grotthuss-type mechanism \cite{30,31} the proton hopping across the hydrogen bond rather than the
transfer due to reorientational motions is the rate-limiting step.

Based on our simulations we postulate the existence of the crossover in the thermal conductivity above 
pressures examined in the experiment. The key to understanding the pressure-driven inverse of the monotonicity 
of the proton conductivity is the behavior of the hydrogen-bond activation barrier with increased pressure.
Experimental verification of our supposition has been scheduled for further studies.

\begin{acknowledgments}
This work was supported by the Polish Ministry of Science and Higher Education through Grant No. N N202 368139.
\end{acknowledgments}


\begin{thebibliography}{99}
\bibitem{1} K.-D.~Kreuer, A.~Fuchs, M.~Ise, M.~Spaeth and J.~Maier, Electrochim.~Acta {\bf 43}, 1281 (1998),
doi:10.1016/S0013-4686(97)10031-7.
\bibitem{2} K.-D.~Kreuer, J.~Membr.~Sci. {\bf 185}, 29 (2001), doi:10.1016/S0376-7388(00)00632-3.
\bibitem{3} M. Schuster, W.H. Meyer, G. Wegner, H.G. Herz, M.~Ise, M.~Schuster, K.D.~Kreuer, J.~Maier, Solid~State~Ion. {\bf 145}, 
85 (2001), doi:10.1016/S0167-2738(01)00917-1.
\bibitem{4} S.R. Benhabbour, R.P. Chapman, G. Scharfenberger, W.H. Meyer, G.R. Goward, Chem. Mater. {\bf 17}, 1605 (2005), 
doi:10.1021/cm048301l.
\bibitem{5} C. Wannek, B. Kohnen, H.-F. Oetjen, H. Lippert and J. Mergel, Fuel Cells {\bf 8}, 87 (2008), doi:10.1002/fuce.200700059.
\bibitem{6} S.U. Celik, A. Aslan, A. Bozkurt, Solid State Ion. {\bf 179}, 683 (2008), doi:10.1016/j.ssi.2008.04.033.
\bibitem{7} J.A. Asensio, E.M. Sanchez, P. Gomez-Romero, Chem. Soc. Rev. {\bf 39}, 3210 (2010), doi:10.1039/B922650H.
\bibitem{8} J.C. MacDonald, P.C. Dorrestein, M.M. Pilley, Crystal Growth \& Design {\bf 1}, 29 (2001), doi:10.1021/cg000008k.
\bibitem{9} S.K. Callear, M.B. Hursthouse and T.L. Threlfall, CrystEngComm {\bf 12}, 898 (2010), doi:10.1039/B917191F.
\bibitem{10} C.~J.~T.~de~Grotthuss, Ann.~Chim.Phys. (Paris) {\bf 58}, 54 (1806).
\bibitem{11} K.D. Kreuer, Solid State Ion. {\bf 136-137}, 149 (2000), doi:10.1016/S0167-2738(00)00301-5.
\bibitem{12} V.V. Sinitsyn, E.G. Ponyatovski, A.I. Baranov, A.V. Tregubchenko, L.A. Shuvalov, Sov. Phys.
JETP {\bf 73}, 386 (1991).
\bibitem{13} V.V. Sinitsyn, A.I. Baranov, E.G. Ponyatovsky, Solid State Ion. {\bf 136-137}, 167 (2000), 
doi:10.1016/S0167-2738(00)00303-9.
\bibitem{14} N.I. Pavlenko, Phys. Rev. {\bf B 61}, 4988 (2000), doi:10.1103/PhysRevB.61.4988.
\bibitem{15} A.I. Baranov, V.V. Grebenev, A.N. Khodan, V.V. Dolinina, E.P. Efremova, Solid State Ion. {\bf 176}, 2871 (2005), 
doi:10.1016/j.ssi.2005.09.018.
\bibitem{16} N. Bagdassarov, G. Lentz, Solid State Commun. {\bf 136}, 16 (2005), doi:10.1016/j.ssc.2005.06.028.
\bibitem{17} P. Ławniczak, M. Zdanowska-Frączek, Z.J. Frączek, K. Pogorzelec-Glaser, Cz. Pawlaczyk, Solid~State~Ion.
{\bf 225}, 268-271 (2012), doi:10.1016/j.ssi.2012.01.041.
\bibitem{18} M. Zdanowska-Frączek, K. Hołderna-Natkaniec, P. Ławniczak, Cz. Pawlaczyk, Solid State Ion. {\bf 237},
40-45 (2013), doi:10.1016/j.ssi.2013.02.013.
\bibitem{19} K. Pogorzelec-Glaser, A. Rachocki, P. Ławniczak, A. Pietraszko. Cz. Pawlaczyk, B. Hilczer,
M. Pugaczowa-Michalska, CrystEngComm {\bf 15}, 1950 (2013), doi:10.1039/C2CE26571K.
\bibitem{20} A. Rachocki, K. Pogorzelec-Glaser, P. Ławniczak, B. Hilczer, A. Łapiński, M. Pugaczowa-Michalska,
M. Matczak, Crystal Growth \& Design {\bf 14}, 1211 (2014), doi:10.1021/cg401742b.
\bibitem{21} S. Summerfield, Philos. Mag. {\bf B52}, 9 (1985), doi:10.1080/13642818508243162.
\bibitem{22} T.~Mas\l{}owski, A.~Drzewi\'{n}ski, J.~Ulner, J.~Wojtkiewicz, M.~Zdanowska-Fr\c{a}czek, K.~Nordlund
and A.~Kuronen, Phys.~Rev. {\bf E 90}, 012135 (2014), doi:10.1103/PhysRevE.90.012135.
\bibitem{23} J.~Hermet, F.~Bottin, G.~Dezanneau, G.~Geneste, Solid~State~Ion. {\bf 252}, 48 (2013), 
doi:10.1016/j.ssi.2013.06.001.
\bibitem{24} K.~A.~Fichthorn and W.~H.~Weinberg, J.~Chem.~Phys. {\bf 95}, 1090 (1991), doi:10.1063/1.461138.
\bibitem{25} A.~Katrusiak, {\it Frontiers of High Pressure Research II: Application of High Pressure to
Low-Dimensional Novel Electronic Materials, NATO Science Series II Vol. 48} (Kluwer Academic Publishers, 2001), pp 73-85.
\bibitem{26} R. Truell, C. Elbaum, B.B. Chick, {\it Ultrasonic methods in solid state physics} (Academic Press,
New York and London, 1969).
\bibitem{27} R.C. Dougherty, J. Chem. Phys. {\bf 109}, 7372 (1998), doi:10.1063/1.477343.
\bibitem{28} H.F. Pollard, {\it Sound waves in solids (Applied Physics Series)} (Law Book Co of Australasia, 1977).
\bibitem{29} A. Katrusiak, J. Mol. Struct. {\bf 474}, 125 (1999), doi:10.1016/S0022-2860(98)00566-3.
\bibitem{30} K.D. Kreuer, A. Fuchs, J. Maier, Solid~State~Ion. {\bf 77}, 157 (1995), doi:10.1016/0167-2738(94)00265-T.
\bibitem{31} R.E. Lechner, Solid State Ion. {\bf 145}, 167 (2001), doi:10.1016/S0167-2738(01)00946-8.
\end{thebibliography}
\end{document}